\documentclass[10pt,journal,compsoc, a4paper]{IEEEtran}

\newcommand{\Supplementary}{Supplementary}

\usepackage[
    colorlinks=true,
    linkcolor=black,
    anchorcolor=black,
    citecolor=black,
    filecolor=black,
    menucolor=black,
    runcolor=black,
    urlcolor=black
]{hyperref}
\usepackage{algorithm}
\usepackage{algpseudocode}
\usepackage{amsmath}
\usepackage{xurl}
\usepackage{mathtools}
\usepackage{cleveref}
\usepackage{orcidlink}
\usepackage{caption}
\usepackage{extarrows}
\usepackage{tabularx}
\usepackage{booktabs}
\usepackage[utf8]{inputenc}
\usepackage{textalpha}
\usepackage{adjustbox}
\usepackage{graphicx}
\usepackage[table,xcdraw]{xcolor}
\usepackage{makecell}
\usepackage{lastpage}
\usepackage{float}
\usepackage{placeins}
\usepackage{tikz}
\usepackage{circuitikz}
\usetikzlibrary {arrows.meta}
\makeatletter
\ctikzset{bipoles/ravenspring/height/.initial=.5}
\ctikzset{bipoles/ravenspring/width/.initial=.5}
\def\pgf@circ@ravenspring@path#1{\pgf@circ@bipole@path{ravenspring}{#1}}
\compattikzset{ravenspring/.style = {\circuitikzbasekey, /tikz/to path=\pgf@circ@ravenspring@path, l=#1}}
\pgfcircdeclarebipole{}
  {\ctikzvalof{bipoles/ravenspring/height}}
  {ravenspring}
  {\ctikzvalof{bipoles/ravenspring/height}}
  {\ctikzvalof{bipoles/ravenspring/width}}
  {%
    \pgfsetlinewidth{\pgfkeysvalueof{/tikz/circuitikz/bipoles/thickness}\pgfstartlinewidth}
    \pgftransformationadjustments

    \pgfmathsetlength{\pgf@circ@res@step}
    {(\ctikzvalof{bipoles/ravenspring/width}*\pgf@circ@Rlen+\pgfhorizontaltransformationadjustment*.5*\pgflinewidth)/16}

    \pgfpathmoveto{\pgfpoint{\pgf@circ@res@left-\pgfhorizontaltransformationadjustment*0.5*\pgflinewidth}{\pgf@circ@res@zero}}
    \pgfsetcornersarced{\pgfpoint{.5\pgf@circ@res@up}{.5\pgf@circ@res@up}}
    \pgf@circ@res@other = \pgf@circ@res@left
    \advance\pgf@circ@res@other by 0.50\pgf@circ@res@step
    \pgfpathlineto{\pgfpoint{\pgf@circ@res@other}{\pgf@circ@res@up}}
    \advance\pgf@circ@res@other by 1.25\pgf@circ@res@step
    \pgfpathlineto{\pgfpoint{\pgf@circ@res@other}{\pgf@circ@res@down}}
    \advance\pgf@circ@res@other by 1.50\pgf@circ@res@step
    \pgfpathlineto{\pgfpoint{\pgf@circ@res@other}{\pgf@circ@res@up}}
    \advance\pgf@circ@res@other by 1.75\pgf@circ@res@step
    \pgfpathlineto{\pgfpoint{\pgf@circ@res@other}{\pgf@circ@res@down}}
    \advance\pgf@circ@res@other by 2.00\pgf@circ@res@step
    \pgfpathlineto{\pgfpoint{\pgf@circ@res@other}{\pgf@circ@res@up}}
    \advance\pgf@circ@res@other by 2.25\pgf@circ@res@step
    \pgfpathlineto{\pgfpoint{\pgf@circ@res@other}{\pgf@circ@res@down}}
    \advance\pgf@circ@res@other by 2.50\pgf@circ@res@step
    \pgfpathlineto{\pgfpoint{\pgf@circ@res@other}{\pgf@circ@res@up}}
    \advance\pgf@circ@res@other by 2.75\pgf@circ@res@step
    \pgfpathlineto{\pgfpoint{\pgf@circ@res@other}{\pgf@circ@res@down}}
    \advance\pgf@circ@res@other by 1.50\pgf@circ@res@step
    \pgfpathlineto{\pgfpoint{\pgf@circ@res@other}{\pgf@circ@res@zero}}
    \pgfsetbuttcap
    \pgfsetbeveljoin
    \pgfusepath{stroke}
  }
\makeatother

\crefname{figure}{Fig.}{Figs.}
\Crefname{figure}{Fig.}{Figs.}
\crefname{table}{Table}{Tables}
\Crefname{table}{Table}{Tables}
\crefname{equation}{Eq.}{Eqs.}
\Crefname{equation}{Eq.}{Eqs.}

\usepackage[style=ieee, mincitenames=1, maxcitenames=2, minbibnames=1, maxbibnames=2, backend=biber, sorting=none, sortcites=true, doi=true, url=false, citestyle=numeric-comp]{biblatex}

\AtEveryBibitem{
    \clearfield{urlyear}
    \clearfield{urlmonth}
    \clearfield{url}
    \clearfield{note}
}

\DefineBibliographyStrings{english}{
  url = \adddot\addspace Available ,
}

\makeatletter
\renewcommand{\abstract}{\normalfont\@IEEEtweakunitybaselinestretch{1.15}\sffamily
    \if@twocolumn
      \@IEEEabskeysecsize\noindent
    \else
      \bgroup\par\addvspace{0.5\baselineskip}\centering\vspace{-1.78ex}\@IEEEabskeysecsize\par\addvspace{0.5\baselineskip}\egroup\quotation\@IEEEabskeysecsize%
    \fi\@IEEEgobbleleadPARNLSP}
\makeatother

\begin{document}

\title{Closed-loop Neuroprosthetic Control through Spared Neural Activity Enables Proportional Foot Movements after Spinal Cord Injury
}

\author{
    Vlad Cnejevici\textsuperscript{1}  \orcidlink{0000-0003-2291-5941},
    Matthias Ponfick\textsuperscript{2} \orcidlink{0000-0002-9944-6007},
    Dietmar Fey\textsuperscript{3}
    \orcidlink{0000-0002-6077-4732},
    Raul C. S\^impetru\textsuperscript{1}  \orcidlink{0000-0003-0455-0168},
    and

    Alessandro Del Vecchio\textsuperscript{1,\;†} \orcidlink{0000-0002-7118-5554}
    \thanks{
        This work was supported by the European Research Council (ERC) under grant 101118089, by the German Federal Ministry of Research, Technology and Space (BMFTR) under grant 16SV9246, by the German Research Foundation (DFG) under grant 523352235, and by the Bavarian Ministry of Economic Affairs, Regional Development and Energy (StMWi) under grants MV-2303-0006 and LSM-2303-0003.
    }
    \thanks{
        \textsuperscript{1} Department of Artificial Intelligence in Biomedical Engineering, Friedrich-Alexander-Universität Erlangen-Nürnberg, 91052 Erlangen, Germany (e-mail: alessandro.del.vecchio@fau.de).
    }
    \thanks{
        \textsuperscript{2} Querschnittzentrum Rummelsberg, Krankenhaus Rummelsberg GmbH, 90592 Schwarzenbruck, Germany (e-mail: Matthias.Ponfick@Sana.de).
    }
    \thanks{
        \textsuperscript{3} Chair of Computer Science 3, Friedrich-Alexander-Universität Erlangen-Nürnberg, 91058 Erlangen, Germany (e-mail: dietmar.fey@fau.de).
    }
    \thanks{\textsuperscript{†} Correspondence to: alessandro.del.vecchio@fau.de}
}

\IEEEtitleabstractindextext{%
    \begin{abstract}
        \normalsize
        Loss of voluntary foot movement after spinal cord injury (SCI) can significantly limit independent mobility and quality of life. To improve motor output after injury, functional electrical stimulation (FES) is used to deliver stimulation pulses through the skin to affected muscles. While commercial FES systems typically use motion-based triggers, prior research shows that spared movement intent can be decoded after SCI using surface electromyography (EMG). Our aim is to assess how well spared neural signals of the lower limb after SCI can be decoded and used to control electrical stimulation for restoring foot movement. We developed a wearable machine learning-powered neuroprosthetic that records EMG from the affected lower limb using a 32-channel electrode bracelet and enables closed-loop control of a FES device for foot movement restoration. Five participants with SCI used the predicted control signal to follow trajectories on a screen with their foot and achieve distinct motor activation patterns for foot flexion, extension, and inversion or eversion. Three of these participants also achieved 2 proportional activation levels during foot flexion/extension with $>$70\% accuracy. To validate how these neural signals can be used for closed-loop neuroprosthetic control, two participants used their decoded activity to control a FES device and stimulate their affected foot. This resulted in an increased foot flexion range for both participants of 33.6\% and 40\% of a functional healthy range, respectively (p $<$ 0.001). One of the participants also achieved voluntary proportional control of up to 6 stimulation levels during foot flexion/extension. These results suggest that wearable EMG decoding coupled with FES systems provides a scalable strategy for closed-loop neuroprosthetic control supporting voluntary foot movement.

    \end{abstract}
}

\maketitle
\IEEEdisplaynontitleabstractindextext
\IEEEpeerreviewmaketitle

\makeatletter
\def\ps@headings{%
    \def\@oddhead{\hbox{}\@IEEEheaderstyle\leftmark\hfil\thepage/\pageref{LastPage}}\relax
    \def\@evenhead{\@IEEEheaderstyle\thepage/\pageref{LastPage}\hfil\leftmark\hbox{}}\relax
    \let\@oddfoot\@empty
    \let\@evenfoot\@empty
}
\def\ps@IEEEtitlepagestyle{%
    \def\@oddhead{\hbox{}\@IEEEheaderstyle\leftmark\hfil\thepage/\pageref{LastPage}}\relax
    \def\@evenhead{\@IEEEheaderstyle\thepage/\pageref{LastPage}\hfil\leftmark\hbox{}}\relax
    \let\@oddfoot\@empty
    \let\@evenfoot\@empty
}
\makeatother
\pagestyle{headings}
\thispagestyle{IEEEtitlepagestyle}

\begin{refsection}

    \IEEEraisesectionheading{\section{Introduction}\label{sec:introduction}}
    \IEEEPARstart{L}{oss} of voluntary foot movement following a neurological injury such as spinal cord injury (SCI) \cite{van_der_salm_gait_2005}, stroke \cite{jakubowitz_behandlungsoptionen_2017}, or peripheral nerve damage \cite{aprile_multicenter_2005} can reduce both mobility and quality of life. One consequence of such conditions is drop foot, characterized by difficulty lifting the foot due to disrupted neural pathways between the brain and leg muscles. A common solution for individuals with drop foot is the use of neuroprosthetic devices based on functional electrical stimulation (FES). These devices deliver stimulation pulses transcutaneously (through the skin) via electrodes placed on the skin surface over the affected muscles \cite{gupta_neuroprosthetics_2023}. By stimulating affected muscles to contract, FES devices can assist users during activities of daily living \cite{liberson_functional_1961, taylor_long-term_2013, everaert_does_2010, kapadia_randomized_2014, field-fote_combined_2001}, as well as improve muscle strength \cite{gupta_neuroprosthetics_2023, hjeltnes_functional_1990} and motor output \cite{ambrosini_cycling_2012}.

    \begin{figure*}[!t]
        \centering
        \includegraphics[width=0.975\textwidth]{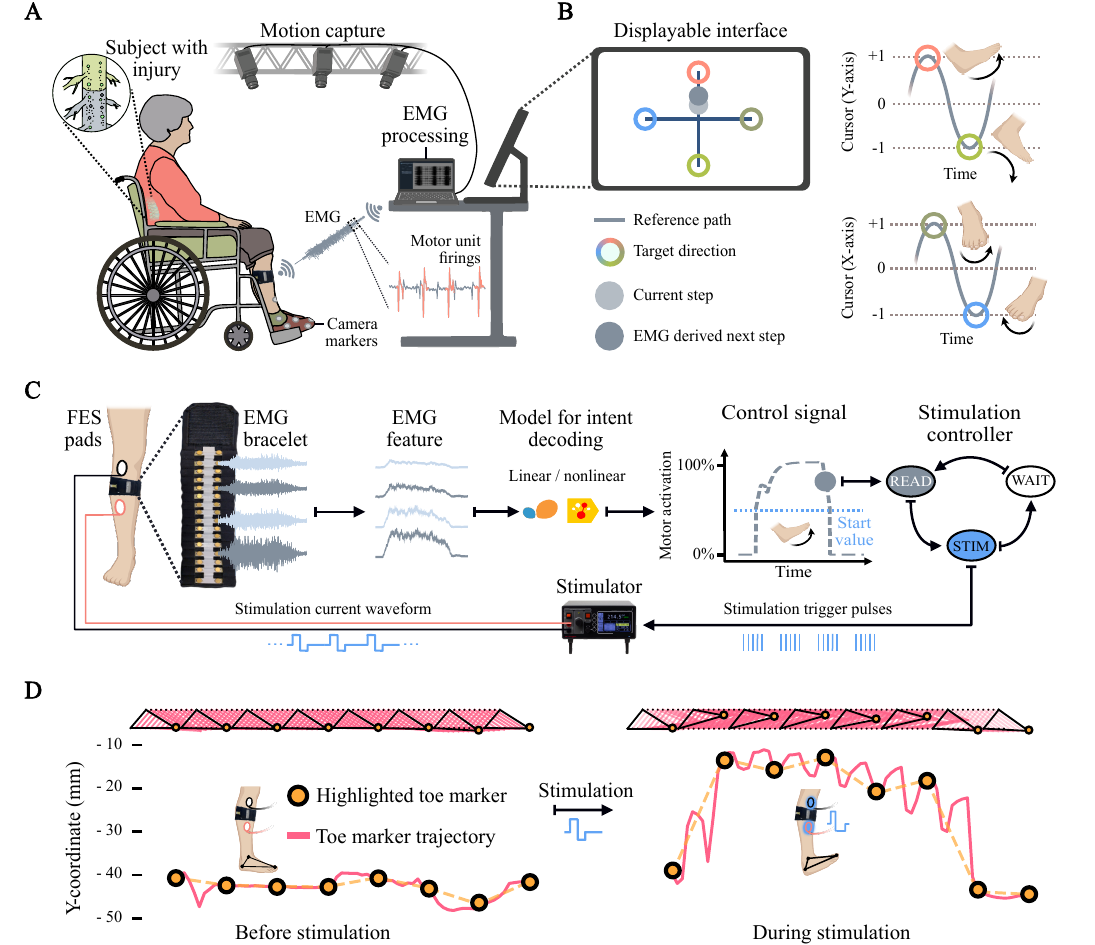}
        \caption{\textbf{Closed-loop controller overview.}
            \textbf{A.} Five individuals with spinal cord injury (S1-S5; \cref{tab:participant_list}) attempted foot movements while spared surface EMG was recorded using a 32-channel electrode bracelet. Two participants (S1-S2) also had their foot and shin kinematics recorded using motion-tracking cameras and infrared reflective markers for the electrical stimulation tests. Example motor unit firings shown for S2. \textbf{B.} A virtual interface displayed a 2D reference cursor as a movement guide. A second cursor provided real-time visual feedback based on the EMG-decoded movement intent. The position of both cursors is normalized between -1 and 1. \textbf{C.} EMG features were extracted and fed to a linear discriminant analysis (linear) and Gradient Boosted Decision Tree (nonlinear) machine learning model, whose predictions updated the cursor position. The predicted cursor also controlled an electrical stimulator in closed-loop, triggering stimulation at a start intensity once the cursor crossed a preset threshold. Stimulation was delivered using a state-machine controller with alternating reading, stimulation, and waiting states, where the stimulation intensity was determined by the cursor value during the reading state. Example cursor signal shown for S5. \textbf{D.} Example toe trajectory in a 2D side projection is shown for S1 before (left) and during stimulation (right).}
        \label{fig:intro_fig}
    \end{figure*}

    Electrical stimulation can be controlled through external trigger events such as button presses or using continuous sensor feedback to detect the user's movement intent. Most commercial \cite{hausdorff_effects_2008, scott_quantification_2013, berenpas_kinematic_2018, embrey_functional_2010} and research systems \cite{heller_automated_2013, muller_adaptive_2020, meng_functional_2017, wiesener_robust_2016} rely on foot switches or motion and tilt sensors to trigger stimulation, which may be limited by the user's pathological range of motion (RoM). Recent works \cite{jung_machine-learning-based_2024, ferrari_emg-triggered_2023, petersen_novel_2020, xu_closedloop_2024} have explored FES control of the lower limb using surface electromyography (EMG), a non-invasive measure of voluntary motor unit (MU) ensemble activity \cite{merletti_electromyography_2004}. These studies, done in both healthy \cite{jung_machine-learning-based_2024, ferrari_emg-triggered_2023, petersen_novel_2020} and stroke populations \cite{xu_closedloop_2024}, have also involved reading EMG from muscles unaffected by the injury to control FES parameters \cite{xu_closedloop_2024}. There remains an opportunity to extend these approaches to individuals with SCI and to stimulation control using EMG recorded from muscles directly affected by the injury.

    In our previous work on the upper limb, we identified spared neural activity associated with hand movement using high-density EMG in both healthy individuals \cite{cakici_generalized_2022, simpetru_learning_2024} and in people with fully paralyzed hands after SCI \cite{oliveira_direct_2024, simpetru_myogestic_2025, ting_sensing_2021}. We have also shown how individuals with SCI can use this activity to accurately control a virtual hand \cite{oliveira_direct_2024, simpetru_myogestic_2025}. Here we adapted this approach to assess the degree of spared neural activity present in the lower limb after SCI. Additionally, we aimed to test if this activity could be decoded from surface high-density EMG using machine learning (ML). We hypothesized that: (1) spared EMG from affected lower limbs contains decodable movement intent, (2) decoded intent can drive an FES system in closed-loop, and (3) FES guided by decoded intent increases functional RoM beyond the pathological movement of the user.

    To test these hypotheses, we developed a neuroprosthetic to decode spared foot movement intent in five motor-incomplete individuals with SCI, all with partial motor function preserved below the injury level (S1-S5; \cref{tab:participant_list}). This cohort included both individuals with chronic SCI ($>$1 year post-injury; S1-S2) as well as acute SCI ($<$1 year post-injury; S3-S5). Spared surface EMG from the affected lower limb was recorded using a wearable high-density electrode bracelet validated in our previous study on the upper limb \cite{simpetru_myogestic_2025} (\cref{fig:intro_fig}A). Participants attempted to follow a trajectory on a screen using an EMG-controlled cursor through the ML model prediction. This prediction provided real-time visual feedback (\cref{fig:intro_fig}B) and controlled an electrical stimulator in closed-loop to restore lost foot movement (\cref{fig:intro_fig}C,D).

    \section{Methods}

    \subsection{Study design}
    We recruited five individuals with SCI (denoted S1-S5; \cref{tab:participant_list}), consisting of 2 male and 3 female participants aged 58.8 $\pm$ 8.4 years. All participants provided written informed consent and the study was conducted in accordance with the Declaration of Helsinki. All procedures were approved by the ethics committees of the Friedrich-Alexander-Universität Erlangen-Nürnberg (applications 24-208-B and 24-286-B approved on 26 June 2024 and 15 August 2024, respectively).

    The experiment protocol consisted of an offline and an online phase. During the offline calibration phase, participants had their spared EMG recorded for model training while they attempted maximal voluntary contractions for each of the following ankle movements: foot flexion/extension (dorsiflexion/plantarflexion) and foot sole inward/outward turn (inversion/eversion). This was followed by an online phase where they used real-time visual feedback provided by the ML model output to follow the trajectories of a 2D cursor on the screen. For both of these phases, two ML models were used for intent decoding: a linear and a nonlinear classifier (\cref{fig:intro_fig}C).

    During online evaluation, participants followed the trajectories of a reference cursor for 3-5 contraction ramps for each evaluated ankle movement while using a second EMG-controlled cursor (\cref{fig:intro_fig}B). This cursor provided real-time visual feedback of the ML decoded movement and motor activation level (\cref{fig:intro_fig}B, C), and further served as a control signal for an electrical stimulator (\cref{fig:intro_fig}C, D). Contraction ramps were performed at 0.1 Hz with a minimum duration of 0.5 s for holding and for resting.

    Minor protocol adjustments were made for S3 and S5 (\cref{tab:participant_list}) to improve comfort and reduce fatigue. For S3, sustained holds during offline calibration were replaced with shorter alternating contraction-rest phases following the same reference pattern as used in the online evaluation. Additionally, only the nonlinear model was used online for S3 because the linear model's predictions were too unstable for further use. For S5, testing was restricted to movements with EMG activity above baseline noise (rest, dorsiflexion and inversion), while hold durations were increased to 2 s to facilitate easier reference cursor tracking.

    \begin{table*}[!t]
        \centering
        \caption{
            \textbf{Participant characteristics.} Age is given in 5-year ranges to protect the participants' privacy. IL = Injury Level. ZPP = Zone of Partial Preservation \cite{rupp_international_2021}. AIS = ASIA Impairment Scale \cite{rupp_international_2021}. DF = Dorsiflexion. I = Inversion. PF = Plantarflexion. Muscle strength was graded using the standardized scale from the Medical Research Council \cite{naqvi_muscle_2025}.
        }
        \label{tab:participant_list}

        \resizebox{\textwidth}{!}{%
            \begin{tabular}{c >{\centering\arraybackslash}p{3.2cm} >{\centering\arraybackslash}p{3.2cm} c c c c c}
                \toprule
                Subject ID & Performed test                                                         & \makecell{Injury type                                                           \\(IL/ZPP/AIS)} & Time since injury & \makecell{Age\\(5-year range)} & Sex & Measured side & \makecell{DF/PF muscle\\grade (0-5)} \\
                \midrule
                \rowcolor{gray!15}
                S1         & Electrical stimulation\newline Task separation\newline Target reaching & Traumatic SCI\newline (L1/S1/D)            & 46 years & 60-64 & M & right & 0/3 \\
                S2         & Electrical stimulation\newline Task separation\newline Target reaching & Spinal cord inflammation\newline (L3/S1/D) & 4 years  & 60-64 & M & left  & 0/1 \\
                \rowcolor{gray!15}
                S3         & Task separation                                                        & Spinal cord inflammation\newline (L2/S1/D) & 3 months & 45-49 & F & left  & 1/3 \\
                S4         & Task separation\newline Target reaching                                & Traumatic SCI\newline (L2/S1/D)            & 1 month  & 55-59 & F & left  & 1/3 \\
                \rowcolor{gray!15}
                S5         & Task separation (DF \& I only)\newline Target reaching (DF only)       & Surgical complication\newline (L1/S1/C)    & 2 months & 65-69 & F & left  & 3/2 \\
                \bottomrule
            \end{tabular}%
        }
    \end{table*}

    \subsection{EMG acquisition}
    The EMG bracelet (HD20MM1602B, OT Bioelettronica S.r.l., Turin, Italy) consisted of 32 dry gold-plated copper electrodes distributed across 2 columns with 16 rows. A wireless probe (Muovi, OT Bioelettronica S.r.l., Turin, Italy) was used to collect EMG signals at 2000 Hz by streaming EMG segments at 111 Hz ($\sim$ every 9 ms). These segments were then concatenated in a real-time buffer to be used for the calculation of the root mean squared (RMS) values. This buffer was set by default to $\sim$252 ms and reduced to $\sim$120 ms for the proportional electrical stimulation test (see Proportional stimulation) to explore if model responsiveness could be increased without a decrease in accuracy. The EMG signals were recorded in monopolar derivation.

    Two EMG bracelet lengths were available (23 and 33 cm) and the one closest to the participant's leg circumference was selected to ensure good electrode-skin contact and minimal motion artefacts. An adhesive reference electrode (Adhesive electrode 4x4 cm, Axion GmbH, Leonberg, Germany) was attached to the medial malleolus of the leg (\cref{fig:sup_marker}A). The EMG signals were bandpass filtered (10-500 Hz) and notch filtered (50 Hz) to remove noise and preserve relevant MU information \cite{merletti_electromyography_2004}, after which RMS features were computed and used as inputs for the ML models (\cref{fig:intro_fig}C).

    \subsection{Model training and optimization}
    To determine the residual ankle movement intent of the participants with SCI, we implemented two ML classifier models (\cref{fig:intro_fig}C): a linear discriminant analysis (LDA) \cite{pedregosa_scikit-learn_2011} and a gradient-boosted decision tree model using the CatBoost implementation \cite{prokhorenkova_catboost_2018, dorogush_catboost_2018}. The LDA provided a linear baseline for evaluating movement separability, while CatBoost could capture nonlinear class boundaries. Additionally, LDA has been used in previous myoelectric decoding studies \cite{al-quraishi_classification_2017, noor_decoding_2021, lyons_case_2015}, while CatBoost has been validated in our recent work \cite{simpetru_myogestic_2025}.

    Each model was trained using a brief offline calibration recording consisting of 10 s of rest followed by a 10 s maximal voluntary contraction for each evaluated ankle movement. Participants practiced each movement for $\sim$5-10 min to confirm EMG activity above baseline and identify a comfortable maximal contraction level.

    Five additional healthy participants (1 female and 4 males; ages 24.8 $\pm$ 2.79 years) were recruited for preliminary EMG recordings to validate model performance and identify optimal parameters for the CatBoost model. These participants performed both the offline calibration as well as the online trajectory following test (see Task separation). The CatBoost hyperparameters were optimized on the offline test data recorded from healthy participants using the Optuna software \cite{akiba_optuna_2019}. The following CatBoost parameters were identified as optimal for our tasks: 300 tree iterations with a tree depth of 10 nodes, a learning rate of 0.04 and an L2 regularization factor of 0.014. For the online test, the healthy participants achieved sufficiently low online error to support further model use (normalized mean absolute error $<$ 0.2 for both models).

    \subsection{Visual interface and cursor control}
    We developed a 2D cursor-based visual interface to guide individuals with SCI in attempting foot movements through a reference cursor moving on the screen (\cref{fig:intro_fig}B). The interface allowed flexible mapping between any cursor movement direction and requested movement label (\Supplementary{} \cref{fig:sup_cursor}A, B). Visually intuitive associations between foot movement and cursor direction were made as follows (\cref{fig:intro_fig}B): up for dorsiflexion, down for plantarflexion, right for inversion, and left for eversion. The reference cursor oscillated linearly along the X- or Y-axis during each movement to indicate the required pace and activation level, similar to prior myocontrol systems \cite{hahne_simultaneous_2018, kamavuako_affordable_2021, nowak_simultaneous_2023}.

    To create a continuous signal from ML model outputs that could be used both as a visual feedback for the decoded intent and to control a FES device, a smoothing exponential decay factor was applied (\cref{fig:sup_cursor}C). Based on the model-predicted movement, a cursor end position would be assigned as a target ($\pm$1.0 on the Y-axis for dorsi-/plantarflexion and $\pm$1.0 on the X-axis for inversion/eversion). After this, the step size required for the cursor to reach that target would be divided by this decay factor. Based on preliminary tests, we found the factor value of 50 (for an EMG streaming rate of 111 Hz) to be suitable for our application.

    \subsection{Kinematics recording}
    Ten infrared cameras (PrimeX 22, OptiTrack, NaturalPoint, Inc., Corvallis, Oregon, USA) were used to record the positions of reflective markers (14 mm, x-base) placed on the foot and shin using Velcro-friendly shoe wraps and skin-adhesive tape. After calibration, the system achieved a tracking error $<$ 0.3 mm, enabling detection of subtle foot orientation changes. Motion was recorded at 120 Hz using Motive 3.0.3 software (OptiTrack, NaturalPoint, Inc., Corvallis, Oregon, USA) and lowpass filtered at 2 Hz. Missing marker data due to occlusion was interpolated or estimated from neighboring markers. Motion data and EMG signals were synchronized via a Python-based graphical user interface automation that triggered both recordings simultaneously \cite{sweigart_pyautogui_2021}.

    \begin{figure*}[!t]
        \centering
        \includegraphics[width=0.975\textwidth]{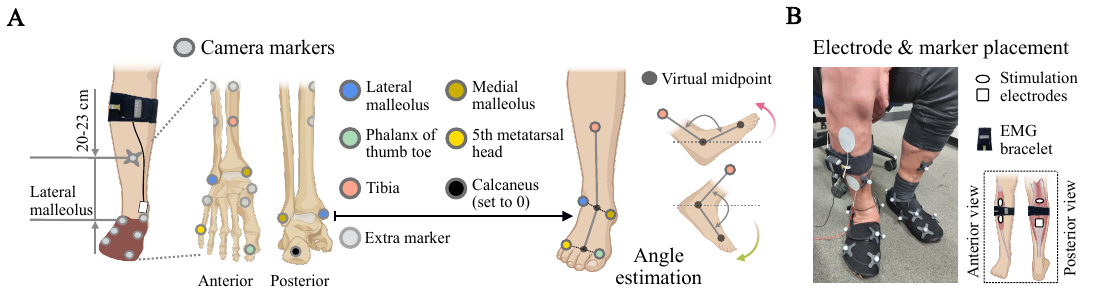}
        \caption{\textbf{Camera marker and electrode placement on the legs.}
            \textbf{A.} The EMG wearable bracelet was placed over the muscle bellies on the upper part of the shin below the knee, while a reference adhesive electrode was placed over the medial malleolus. Additionally, infrared markers on the shin and foot were used to compute 3D coordinates and joint angles for assessing the ankle range of motion. The primary camera markers, shown in color, enabled dorsi-/plantarflexion angle measurement, while the calcaneus marker corrected for the initial foot position offset. Marker placement was chosen to avoid interference with the EMG bracelet, and with the reference electrode on the medial malleolus. Additional markers helped reconstruct the key positions of the primary markers in case of occlusion. \textbf{B.} The EMG bracelet was placed below the knee over the muscle bellies of the measured leg, while stimulation electrodes were positioned above and below the bracelet on the anterior and posterior sides of the same leg. Infrared camera markers were placed on both legs to capture the participant's ankle range of motion before and during stimulation.
        }
        \label{fig:sup_marker}
    \end{figure*}

    Foot motion was tracked per leg using five markers placed on anatomical landmarks of the foot \cite{leardini_rear-foot_2007} and one on the tibia, which was sufficient to compute dorsi-/plantarflexion angles. Up to ten additional markers per leg were added on the tibia and foot to aid marker position reconstruction during marker occlusions. The RoM was calculated as the difference between the joint angle at each time point and the angle at the start of each kinematic recording, so that only motion during the contraction ramps was evaluated. The baseline shift between the joint angle values at the beginning and end of each ramp was also removed so as not to influence RoM calculation, as this shift could simply indicate that the participant reached an alternative resting position to the starting one. Marker placement and ankle angle calculations are shown in \cref{fig:sup_marker}A.

    For this work, we compared achieved RoM values with typical RoM (tRoM) joint angle values reported for gait recordings of older healthy populations \cite{silder_active_2008, jiang_gait_2025, johnson_three-dimensional_2021, zhong_gait_2021}. From the reported values, we selected 20$^\circ$ for dorsiflexion and 16$^\circ$ for plantarflexion as upper-range reference values. To confirm whether these values were appropriate references for this study, we recorded the maximum RoM achieved at comfortable effort with the healthy leg for S1-S2. The maximum dorsiflexion RoM for the healthy foot was 11.39 and 20.6$^\circ$ for S1-S2, respectively, while for plantarflexion it was 9.97$^\circ$ for S1. This showed that the selected reference values were relevant for the recorded participants.

    Because not all participants could be recorded in a laboratory equipped with a motion-capture camera system, we evaluated the pathological RoM for all participants with SCI using markerless pose estimation. Residual foot RoM was recorded using a webcam and foot kinematics were extracted with DeepLabCut (version 3.0.0rc9) \cite{mathis_deeplabcut_2018, nath_using_2019}. The tracked points on the foot and shin were selected to closely match marker locations used during the motion-capture electrical stimulation experiment (\cref{fig:sup_marker}A).

    To correct for perspective distortion from single camera recording, three additional reference points were marked on a rectangular frame parallel to the foot motion plane (e.g., the experiment chair frame) and used with an affine rectification to correct the computed ankle joint angle. Three anatomical landmarks were labeled to estimate dorsi-/plantarflexion angles: the head of the second metatarsal, the midpoint between the lateral and medial malleoli, and a tibial point aligned with the EMG bracelet.

    \subsection{Electrical stimulation}
    For the electrical stimulation tests, we compared affected and unaffected foot kinematics of participants S1 and S2 before and during stimulation. To determine which side was affected or unaffected by the injury, we used the muscle strength scale provided by the Medical Research Council \cite{naqvi_muscle_2025}, where a side was considered affected if it was graded $\leq$ 3 and unaffected if graded $>$ 3. All MRC scores were assessed by a clinician prior to any EMG recording and are reported in \cref{tab:participant_list}.

    For precise tracking of small lower limb movements during stimulation, we used the motion-capture system with infrared cameras to track markers placed on the lower limb (\cref{fig:intro_fig}A and \cref{fig:sup_marker}A; see Kinematics recording). Requested movements were performed bilaterally to compare the affected and unaffected limb RoM before and during stimulation.

    We tested our stimulation controller on two participants (S1-S2), both previous FES orthosis users (S1: L300 Go, Bioness Medical, Inc., Valencia, California, USA; S2: ALFESS, Allard International, Helsingborg, Sweden). Participants underwent a stimulation calibration procedure to identify their maximum comfortable stimulation current (\Supplementary{} \cref{tab:sup_stim_params}) and electrode placement.

    The stimulation controller integrated into the cursor interface (\Supplementary{} \cref{fig:sup_cursor}D) operated as a three-state machine (reading, stimulation, and waiting), governing stimulation and cursor behavior (\cref{fig:intro_fig}C and \cref{fig:electrical_stimulation}A). The predicted cursor controlled the stimulation current proportionally by mapping the Y-axis cursor task range to the stimulation current range from 0 to the user maximum value selected during the stimulation calibration.

    Functional electrical stimulation was delivered to the shin muscles where EMG was recorded using a biphasic constant current stimulator (DS8R, Digitimer Ltd, Welwyn Garden City, United Kingdom) and a D188 remote electrode selector (Digitimer Ltd, Welwyn Garden City, United Kingdom). The device allows precise control of stimulation current (0.1 mA steps) and pulse width (10 µs steps), includes a 300 mJ pulse energy safety limit, and is suitable for benchtop use ($\sim$2.1 kg, 270 x 110 x 225 mm). Adhesive hydrogel electrodes (ValuTrode Cloth Electrodes, Axelgaard Manufacturing Co. Ltd., Fallbrook, California, USA) were placed above and below the 32-channel EMG bracelet: 4x6 cm oval-shaped electrodes or 5x5 cm square-shaped ones over the tibialis anterior muscle for dorsiflexion and the triceps surae muscles for plantarflexion (\cref{fig:intro_fig}C, \cref{fig:sup_marker}B).

    Maximum user pulse amplitude and pulse width were set on the stimulator front panel, while pulse train parameters were configured in the cursor interface (\cref{fig:sup_cursor}). A 300 $\mu$s pulse width and biphasic symmetrical waveform were used to minimize muscle fatigue, reduce the likelihood of muscle spasms due to spasticity \cite{bekhet_effects_2019, kesar_effect_2006} and prevent tissue damage due to stimulation \cite{marquez-chin_functional_2020}. The current used for stimulation was individually adjusted to produce visible, painless contractions (\Supplementary{} \cref{tab:sup_stim_params}), remaining within typical value ranges used for transcutaneous electrical stimulation \cite{marquez-chin_functional_2020}, as well as ranges observed in other spinal cord injury studies \cite{sanna_evaluating_2025, dolbow_body_2014}.

    Stimulation was triggered by a square-wave signal generated from a Raspberry Pi 5 (Raspberry Pi Foundation, Cambridge, United Kingdom). A Transmission Control Protocol connection transmitted pulse parameters and real-time cursor predictions between the host and Pi device. The trigger signal used 5 $\mu$s pulse duration and 35 Hz initial pulse frequency, suitable for tetanic contractions \cite{marquez-chin_functional_2020}, which was later lowered as needed if transmission delays occurred during the experiment (\Supplementary{} \cref{tab:sup_stim_params}).

    To reduce the impact of stimulation artefacts on the EMG signal and the RMS features used for model inputs, a blanking procedure was applied prior to RMS calculation. Whenever a predefined threshold (set to 200 analog-to-digital units) was exceeded in at least half of the EMG channels, a blanking window of 7.5 ms before and 15 ms after the detected artefact was applied. This threshold value was selected from preliminary tests to reliably detect stimulation artefacts across participants. In addition, an exponential moving-average baseline removal filter ($\alpha$ = 0.1) was used to suppress slow signal discharge caused by amplifier saturation and current spread through the skin.

    A graphical user interface tab integrated within our virtual interface enabled real-time adjustment of stimulation pulse train parameters to fine-tune the stimulation control for each participant (\Supplementary{} \cref{fig:sup_cursor}). These included the stimulation start value, pulse frequency, stimulation controller speed, and stimulation and waiting state durations. The controller speed value (1-10) was set to 8 and determined how many iterations the controller waited before updating the cursor position in the reading state. A multiplying factor of 1.0 - (controller speed - 1) × 0.1 (bounded between 0.1 and 1) was applied to the smoothing coefficient to allow a larger initial cursor update at the start of the reading state and compensate for the waiting periods. This position update speed was then exponentially reduced over remaining iterations ($=$ controller speed value) until the cursor position update factor returned to the initial step size.

    \subsection{EMG decomposition}
    To confirm the presence of spared motor unit activity underlying the EMG signals decoded by the ML models, we performed EMG decomposition on a subset of participants (S1-S2). Surface EMG signals used for decomposition were initially bandpass filtered between 20 and 500 Hz using a second order Butterworth filter. To identify individual motor units, a blind source separation method known as the convolution kernel compensation algorithm was applied. This analysis was performed using the DEMUSE software (version 4.5; University of Maribor, Slovenia) for automatic detection of discharge times of the motor units \cite{holobar_multichannel_2007, davies_gradient_2007}. The identified motor unit spike trains were subsequently visually inspected to correct any false positive or false negative detections. From those, only motor units with a pulse-to-noise ratio greater than 35 dB after manual inspection were retained. Afterwards, MU discharge rates were smoothed using a 400 ms Hann window \cite{negro_fluctuations_2009} and correlated with the reference signal.

    \subsection{Evaluation tests}
    To evaluate the decoding and stimulation system, we designed four tests: two electrical stimulation tests (proportional and sustained) to evaluate closed-loop stimulation performance, a task separation test to quantify online cursor control accuracy across four ankle movements, and a target reaching test to assess proportional EMG control at multiple activation levels.

    \subsubsection{Electrical stimulation tests}
    To validate our system's ability to control an assistive device in closed-loop and increase foot RoM after drop foot, we tested in S1 and S2 (\cref{tab:participant_list}) whether the predicted cursor signal could drive electrical stimulation. Both ML models were evaluated prior to applying stimulation and only the model that the participant found easier to control was used further (\Supplementary{} \cref{tab:sup_stim_params}).

    For the proportional control test, S1 performed 3 dorsiflexion and 3 plantarflexion contraction ramps with stimulation to verify reference tracking with the affected foot. This was followed by a randomized sequence of 10 dorsi-/plantarflexion ramps cued verbally (\cref{fig:electrical_stimulation}B) and done in the following order: 2 dorsiflexion, 1 plantarflexion, 4 dorsiflexion, and 3 plantarflexion. Ramps were performed at 0.1 Hz with 0.5 s holding duration. The predicted cursor simultaneously provided visual feedback and drove the stimulation state machine, setting intensity as a percentage of the calibrated maximum (\cref{fig:electrical_stimulation}A; \Supplementary{} \cref{tab:sup_stim_params}).

    For the sustained contraction test, we assessed whether closed-loop stimulation could increase ankle RoM at maximum calibrated intensity compared to the user pathological RoM (\Supplementary{} \cref{tab:sup_stim_params}) and whether the stimulated movement was clearly separable from rest over a prolonged duration. For each participant, we recorded three 10 s contraction ramps with 10 s rest intervals before stimulation, and repeated the same protocol during stimulation. RoM was compared across four conditions: full contraction and rest, each before and during stimulation.

    \subsubsection{Task separation}
    Five participants with SCI (S1-S5; \cref{tab:participant_list}) used their spared EMG to control a 2D cursor and followed the trajectory of a visual reference while attempting four ankle movements (\cref{fig:intro_fig}B): dorsiflexion, plantarflexion, inversion, and eversion. The intersection of the X- and Y- axes was mapped to the resting state, whereas each axis endpoint represented full activation of one ankle movement (\cref{fig:intro_fig}B, \cref{fig:emg_vs_kinematics}A, \cref{fig:electrical_stimulation}A). The recorded EMG signals were decoded into movement intent using ML models (\cref{fig:task_separation}A), while the model output was encoded into a second EMG-controlled cursor serving as real-time feedback on the movement intent. This closed-loop setup enabled us to quantify participant performance with visual feedback and to assess overall model performance (\cref{fig:task_separation}B-D).

    Model performance was evaluated using both offline test data and online cursor-control data. The offline data was split for each recorded movement into training (first and last 40\%) and test data (middle 20\%). Online performance was quantified using the normalized mean absolute error (nMAE) between the EMG-controlled cursor and the reference cursor (\cref{fig:task_separation}B). Decoding accuracy was computed by labeling the reference signal as dorsiflexion, plantarflexion, inversion or eversion if the cursor position on any axis was $>$0.5 or $<$-0.5, and as rest otherwise.

    In a separate analysis to confirm the presence of spared neural activity in the affected limb, we recorded an additional 30 s trial per task in S1-S2 for dorsi-/plantarflexion. Participants followed the same contraction-rest protocol as for the online evaluation. We then decomposed the recorded EMG into MU activity (see EMG decomposition). The identified MUs were considered task-modulated if they showed a correlation coefficient of above 0.7 and a phase-shift variance below 0.7 between their discharge rates and the reference cursor signal, similar to the approach from our previous work \cite{oliveira_discharge_2025}.

    \subsubsection{Target reaching}
    Four participants with SCI (S1, S2, S4 and S5; \cref{tab:participant_list}) attempted to hold dorsi-/plantarflexion at two activation levels by maintaining the predicted cursor within a rectangular target frame displayed in the visual interface (\cref{fig:target_reaching}A). The aim was to assess the participants' intrinsic proportional control by reaching and holding both partial and full EMG activation. For this test, ML models were trained only on rest and dorsi-/plantarflexion data. Cursor smoothing produced intermediate positions during partial EMG activation, which we hypothesized would allow participants to achieve at least a second contraction level with the cursor feedback. The cursor axis was divided into three regions for rest, partial, and full activation (\cref{fig:target_reaching}A). Participants followed two-stage trapezoidal ramps with 10 s at each level.

    Task accuracy was calculated as the proportion of cursor positions within the designated task zone. To assess whether cursor holding reflected true proportional EMG modulation, we also computed the Jensen-Shannon divergence (JSD) between EMG RMS distributions as a model-independent measure of motor activation separability (\cref{fig:target_reaching}A) \cite{lin_divergence_1991}.

    \subsection{Statistical analysis}
    For task separation, a Kruskal-Wallis test ($\alpha$ = 0.05, one-way, unpaired) was applied to nMAE values across tasks (\cref{fig:task_separation}D). A Mann-Whitney U test ($\alpha$ = 0.05, unpaired, two-sided) compared error values between LDA and CatBoost (\cref{fig:task_separation}C). For partial EMG activation during the target reaching test, a Kruskal-Wallis test ($\alpha$ = 0.05) followed by post-hoc two-sided Mann-Whitney U tests with Bonferroni correction was used (\cref{fig:target_reaching}D). A Wilcoxon signed-rank test (paired; $\alpha$ = 0.05) assessed differences between the participant's spared foot kinematics and the neurally driven 2D control signal (\cref{fig:emg_vs_kinematics}B). These nonparametric tests were selected due to small sample sizes per group, making normality assumptions unsuitable.

    For electrical stimulation, a one-way ANOVA ($\alpha$ = 0.05) compared RoM across four conditions (full contraction and rest before/during stimulation), followed by post-hoc independent Welch t-tests with Bonferroni correction (\cref{fig:electrical_stimulation}E). Pearson correlation analysis ($\alpha$ = 0.05) assessed relationships between task accuracy and JSD during partial dorsi-/plantarflexion (\cref{fig:target_reaching}C) and between affected and unaffected ankle RoM before/during stimulation (\cref{fig:electrical_stimulation}B, D; see Proportional stimulation).

    \section{Results}

    \subsection{Spared neural activity and movement intent}
\begin{figure}[!t]
        \centering
        \includegraphics[width=1\columnwidth]{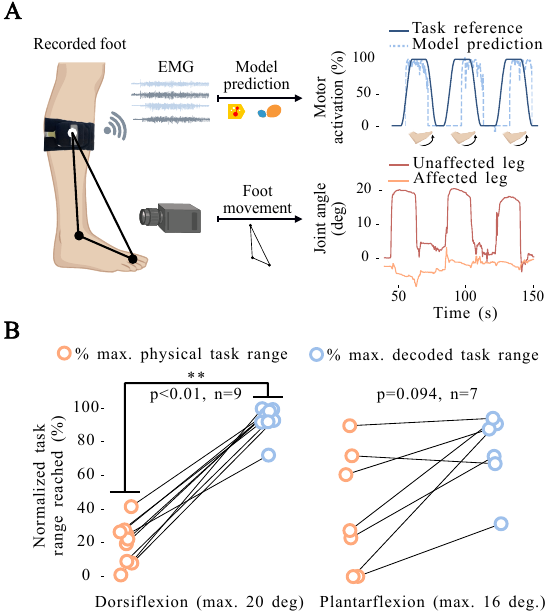}
        \caption{\textbf{Spared neural and physical output after injury.}
            \textbf{A.} Individuals with spinal cord injury (S1-S5; \cref{tab:participant_list}) attempted dorsi-/plantarflexion while residual EMG and foot movement were recorded using a wearable electrode bracelet and webcam. Foot kinematics were extracted using markerless pose estimation, while spared EMG was decoded using a machine learning model. Example intent prediction and kinematics shown for S2. \textbf{B.} Normalized physical task range of the affected foot with respect to typical range of motion for both models is shown during the holding phase (mean computed over middle 3 s). Normalization is used to assess how closely the pathological range of motion and decoded movement intent each approach a healthy motor output. Alongside it, the decoded task range from the predicted cursor signal is shown normalized to the task maximum value of 1.0. Statistical test via Wilcoxon signed-rank test ($\alpha$ $=$ 0.05; p values and sample size shown).}
        \label{fig:emg_vs_kinematics}
    \end{figure}

    We evaluated the stability of ML-based decoding for dorsi-/plantarflexion intent and assessed how well participants could use the decoded cursor control signal to reach their available motor output range (\cref{fig:emg_vs_kinematics}A, B). During these online tests, pathological foot RoM was recorded using a webcam and extracted using DeepLabCut \cite{mathis_deeplabcut_2018, nath_using_2019} (see Methods).

    The RoM achieved during the holding phase of contraction ramps was normalized to the tRoM reference values of 20$^\circ$ for dorsiflexion and 16$^\circ$ for plantarflexion, while the predicted control signal was normalized to the reference cursor maximum. We then assessed how closely each approached a healthy motor output (\cref{fig:emg_vs_kinematics}A, B). Our results show that the decoded neural signal reached a significantly larger proportion of its task maximum than the physical RoM reached the typical healthy range (p $<$ 0.01; \cref{fig:emg_vs_kinematics}B).

    \subsection{Electrical stimulation}

    \begin{figure*}[!t]
        \centering
        \includegraphics[width=0.975\textwidth]{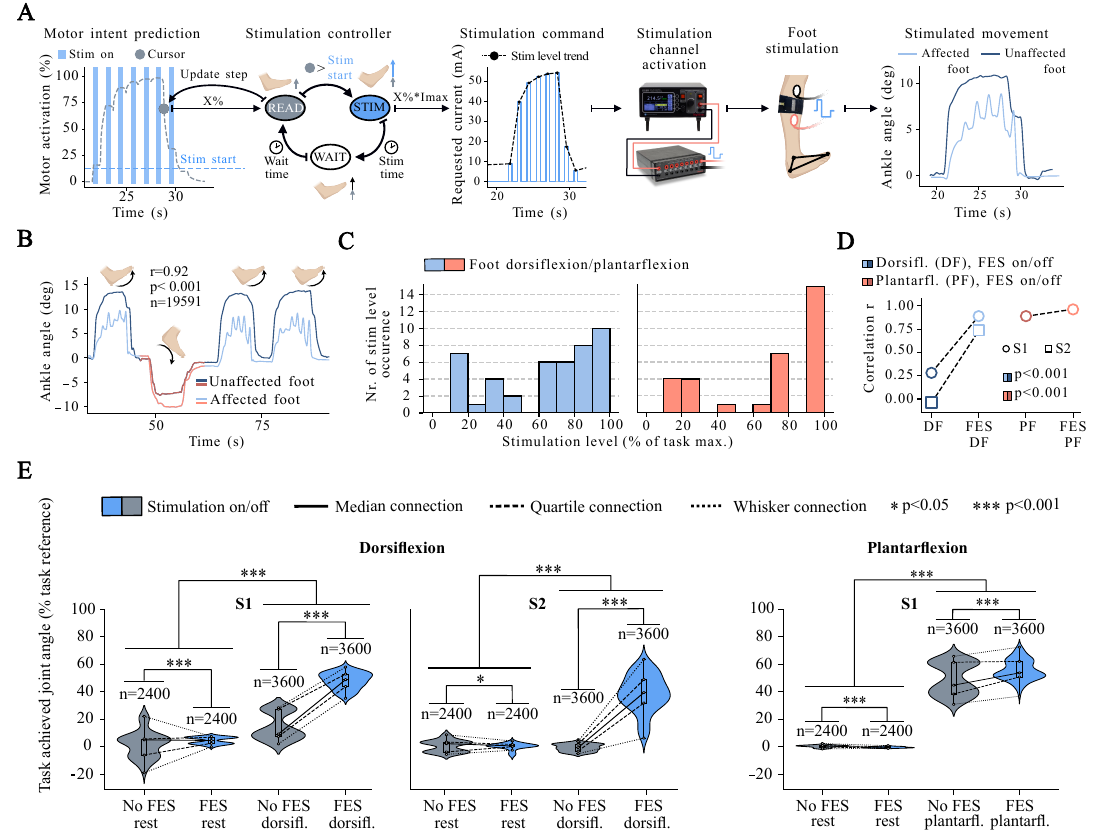}
        \caption{\textbf{Electrical stimulation tests overview.}
            \textbf{A.} Stimulation control pipeline for participants 1 and 2 (S1-S2; \cref{tab:participant_list}). Based on its position on the Y-axis, the predicted cursor proportionally set the stimulation amplitude as a percentage of the maximum current required for the participant during a task. The cursor Y-axis position also selected the active stimulation channel (dorsiflexion for positive values, plantarflexion for negative). When the cursor position crossed a preset threshold, a state-machine controller transitioned from predicted cursor reading to stimulation and waiting states before returning to reading. The stimulation and waiting states transitioned automatically after a time preset by the experimenter. Example predicted cursor, stimulation current, and joint angles for both feet are shown for S1 during a dorsiflexion contraction ramp. \textbf{B.} Example randomized dorsi-/plantarflexion ramps are shown for S1 (correlation coefficient with p-values and sample size included). \textbf{C.} Stimulation levels for S1 binned in 10\% increments of the maximum stimulation current during dorsi-/plantarflexion. We only considered the stimulation levels that were achieved by S1 at least 3 times as consistent. \textbf{D.} Average Pearson correlation coefficient between affected and unaffected foot ramps for S1-S2 before and during stimulation (p-values and sample size shown). \textbf{E.} Task achieved joint angles for S1-S2, normalized to the typical range of motion (20 and 16$^\circ$ for dorsi-/plantarflexion) before and during stimulation. The violin plots and inner box plots show the distribution of achieved range of motion for the middle 10 s of all the performed contractions. Statistical analysis by one-way ANOVA ($\alpha$ $=$ 0.05), followed by post hoc independent Welch's t-tests with Bonferroni correction. *p $<$ 0.05, ***p $<$ 0.001 (p-values and sample size shown).}
        \label{fig:electrical_stimulation}
    \end{figure*}

    \subsubsection{Proportional stimulation}
    We evaluated whether decoded intent from S1 could support graded stimulation during dorsi-/plantarflexion (\cref{fig:electrical_stimulation}A-C). Results showed a significant Pearson correlation between affected and unaffected foot kinematics during dorsi-/plantarflexion (r = 0.92, p $<$ 0.001, n=19591, \cref{fig:electrical_stimulation}B). Additionally, S1 achieved multiple stable stimulation intensity levels (stable if $\geq$ 3 occurrences) during the randomized sequence, including 6 levels for dorsiflexion and 4 for plantarflexion (\cref{fig:electrical_stimulation}C). The stimulation amplitude change over time also tracked the requested reference trajectory (r = 0.69, p $<$ 0.001, n = 76; \cref{fig:electrical_stimulation}A).
    \subsubsection{Sustained stimulation}
    We computed the Pearson correlation coefficient between affected and unaffected foot kinematics. For both participants, correlation increased from r $<$ 0.3 before stimulation to r $>$ 0.7 during stimulated dorsiflexion (p $<$ 0.001; \cref{fig:electrical_stimulation}D). We also quantified RoM changes relative to the selected tRoM values of 20$^\circ$ and 16$^\circ$ for dorsi-/plantarflexion (see Kinematics recording). Results showed a significant dorsiflexion increase for both participants: 33.6\% tRoM for S1 and 40\% tRoM for S2 during stimulation (p $<$ 0.001; \cref{fig:electrical_stimulation}E). Meanwhile for plantarflexion, S1 showed an increase of 6.63\% tRoM during stimulation (p $<$ 0.001; \cref{fig:electrical_stimulation}E).

    \subsection{Task separation}

    \begin{figure*}[!t]
        \centering
        \includegraphics[width=0.975\textwidth]{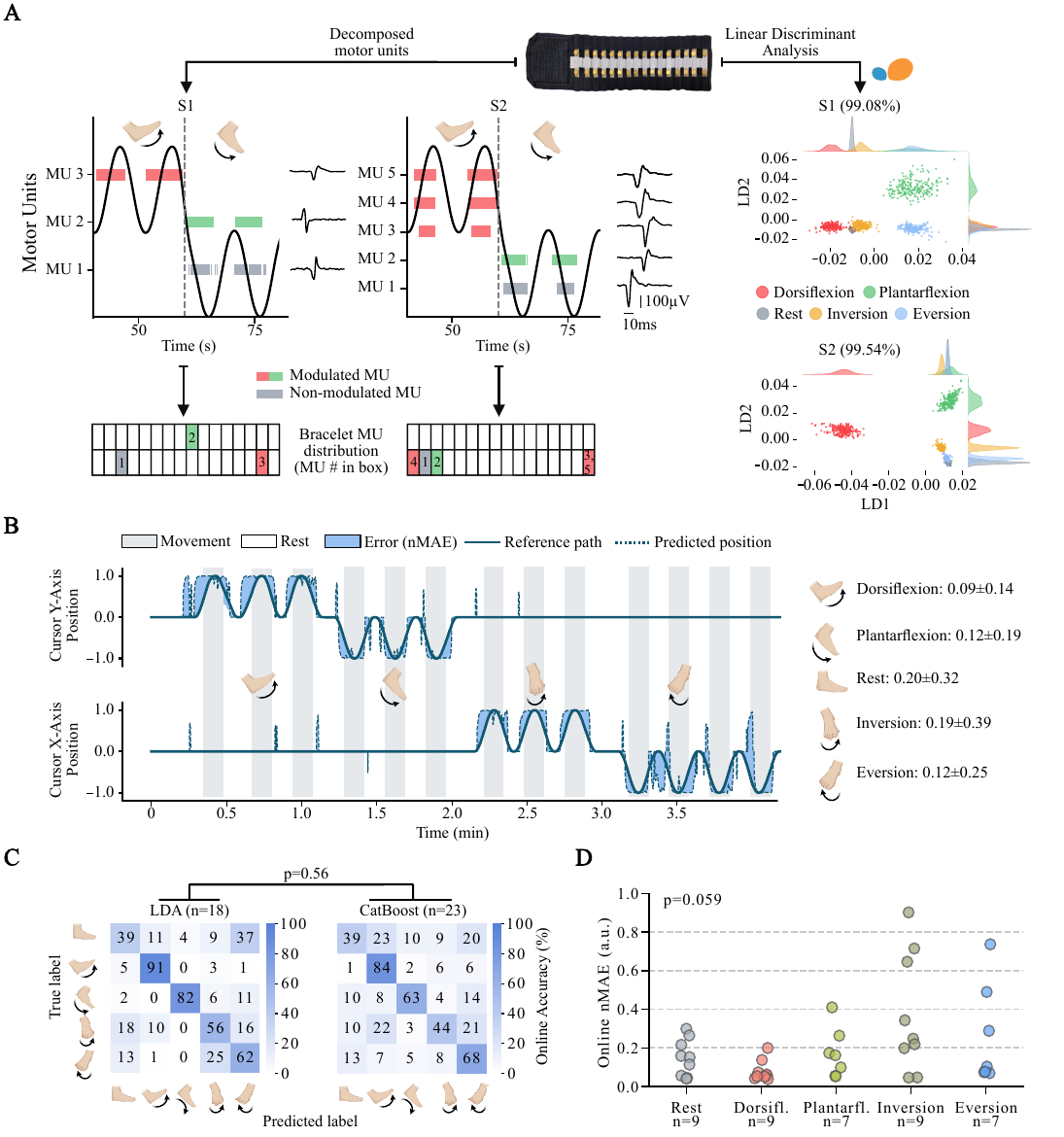}
        \caption{\textbf{Task separation test overview.}
            \textbf{A.} Spared EMG recorded from participants with spinal cord injury (S1-S5; \cref{tab:participant_list}) were decomposed into motor unit (MU) firings (left) and fed to the Linear Discriminant Analysis (LDA) and CatBoost gradient boosted tree models (right). \textbf{Left:} The task-modulated and non-modulated MU firings are shown for S1-S2, along with MU spatial distribution over the EMG bracelet and action potential shapes obtained through 60 ms spike triggered averaging. \textbf{Right}: First two LDA components with kernel density estimates and offline performance for S1-S2 (S3-S5 shown in \Supplementary{} \cref{fig:sup_lda}). LDA components were aligned across participants using Procrustes analysis \cite{goodall_procrustes_1991}. \textbf{B.} Online cursor control performance quantified by normalized mean absolute error (nMAE). Example reference and predicted trajectories with error shown for S4. \textbf{C.} Online model decoding accuracy across participants. The reference cursor was labeled as dorsiflexion, plantarflexion, inversion or eversion if its position on any axis was $>$0.5 or $<$-0.5, and as rest otherwise. Statistical test via Mann-Whitney U test ($\alpha$ $=$ 0.05; p-value and sample size shown). \textbf{D.} Online task performance across all subjects and models. Statistical test via Kruskal-Wallis test ($\alpha$ $=$ 0.05; p-value and sample size shown).}
        \label{fig:task_separation}
    \end{figure*}

    Distinct task-modulated MUs were observed for both dorsiflexion and plantarflexion (\cref{fig:task_separation}A). We identified 2 task-modulated MUs for S1 (1 dorsiflexion, 1 plantarflexion) and 4 for S2 (3 dorsiflexion, 1 plantarflexion).

    Offline decoding accuracy on the test data per subject is shown in \cref{fig:task_separation}A and \Supplementary{} \cref{fig:sup_lda}. Across participants with SCI, task-specific performance showed a mean nMAE $<$ 0.4 for rest, dorsiflexion, and plantarflexion (\cref{fig:task_separation}B, D), with group means of 0.19 $\pm$ 0.20 for LDA and 0.22 $\pm$ 0.22 for CatBoost. Model accuracies across subjects are shown in \cref{fig:task_separation}C, and task-specific nMAE values are shown in \cref{fig:task_separation}D.

    \subsection{Target Reaching}

    \begin{figure*}[!t]
        \centering
        \includegraphics[width=0.975\textwidth]{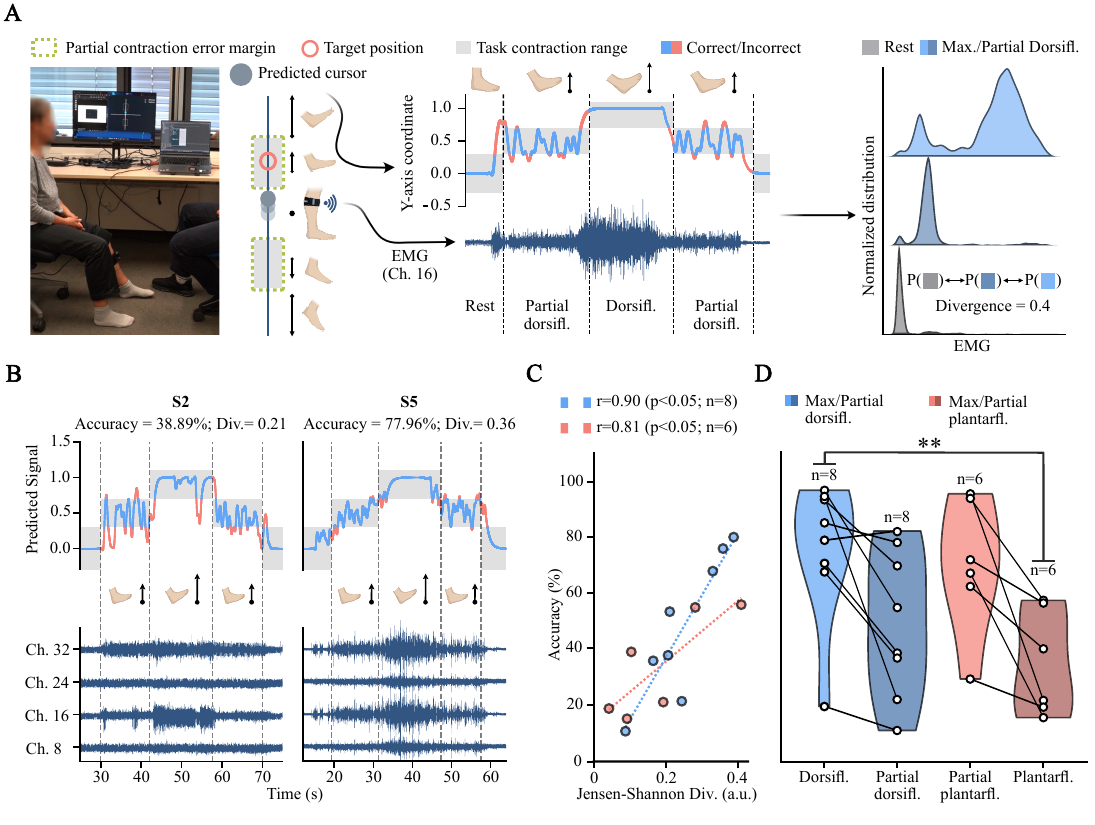}
        \caption{\textbf{Target reaching test overview.}
            \textbf{A.} Four participants with spinal cord injury (S1, S2, S4, S5; \cref{tab:participant_list}) attempted to hold the predicted cursor within the middle $\pm$ 20\% of the task-relevant axis. Accuracy was computed from cursor positions inside and outside the target ranges, while the Jensen-Shannon divergence quantified distinctiveness of EMG root mean squared value distributions (example shown for channel 16 recorded from S1). \textbf{B.} Example predicted signals during full and partial dorsiflexion for S2 and S5, including task accuracy, Jensen-Shannon divergence (Div.), predicted Y-axis trajectories with target ranges, and four representative EMG channels. \textbf{C.} Correlation between accuracy and Jensen-Shannon divergence for partial dorsi-/plantarflexion across subjects and models. Statistical test using the Pearson correlation coefficient ($\alpha$ $=$ 0.05; p-value and sample size shown). \textbf{D.} Task accuracy across all subjects and models. White dots indicate individual subject-model accuracies. Statistical test via Kruskal-Wallis test ($\alpha$ $=$ 0.05; sample size shown), followed by post hoc two-sided Mann-Whitney U tests with Bonferroni correction. **p $<$ 0.01.
        }
        \label{fig:target_reaching}
    \end{figure*}

    Visualization of the predicted cursor signal with a subset of EMG channels is shown for S1 in \cref{fig:target_reaching}A and for S2 and S5 in \cref{fig:target_reaching}B. The JSD and the overall task accuracy are shown in \cref{fig:target_reaching}C, D. Mean task accuracies were 75.70\% $\pm$ 23.46\% for dorsiflexion, 49.18\% $\pm$ 24.49\% for partial dorsiflexion, 69.92\% $\pm$ 22.05\% for plantarflexion, and 35.13\% $\pm$ 17.21\% for partial plantarflexion (\cref{fig:target_reaching}D). Additionally, accuracy per subject and their best performing model are reported in \cref{tab:sup_target_reaching}.

    \begin{table*}[!htbp]
        \centering
        \caption{
            \textbf{Task accuracy in target reaching test for each subject and ML model.} Best performing model for each subject is highlighted in bold.
        }
        \label{tab:sup_target_reaching}

        \newcommand{\best}[1]{\textbf{\boldmath$#1$}}%
        \rowcolors{2}{gray!15}{white}
        \resizebox{\textwidth}{!}{%
            \begin{tabular}{c c c c c c c}
                \toprule
                Subject ID & Overall accuracy (\%) & \multicolumn{2}{c}{Dorsiflexion accuracy (\%)} & \multicolumn{2}{c}{Plantarflexion accuracy (\%)} & Rest accuracy (\%)                                                 \\
                \cmidrule(lr){3-4}\cmidrule(lr){5-6}
                           & LDA (CatBoost)        & Full activation                                & Partial activation                               & Full activation      & Partial activation   &                      \\
                \midrule
                S1         & \best{78.45} (44.61)  & \best{93.13} (96.50)                           & \best{69.64} (54.81)                             & \best{93.69} (95.25) & \best{57.38} (21.78) & \best{94.14} (26.70) \\
                S2         & 37.33 (\best{70.13})  & 19.73 (\best{70.64})                           & 11.10 (\best{38.89})                             & 62.36 (\best{71.80}) & 40.14 (\best{56.42}) & 47.65 (\best{98.67}) \\
                S4         & \best{53.59} (46.12)  & \best{67.58} (94.32)                           & \best{36.86} (22.11)                             & \best{67.06} (29.39) & \best{15.67} (19.40) & \best{77.55} (70.36) \\
                S5         & 84.61 (\best{84.82})  & 78.73 (\best{84.93})                           & 82.06 (\best{77.96})                             & N/A                  & N/A                  & 92.92 (\best{91.97}) \\
                \bottomrule
            \end{tabular}%
        }
    \end{table*}

    \FloatBarrier
    \section{Discussion}
    In our study, we developed a neuroprosthetic that decodes spared lower limb EMG in real-time in individuals with SCI (\cref{fig:intro_fig}A, C) to identify both the spared movements and the potential spared proportional control in foot activation. Our system also supports proportional foot movement after drop foot using closed-loop electrical stimulation (\cref{fig:intro_fig}C, D, and \cref{fig:electrical_stimulation}A-E).

    Across five participants with SCI (S1-S5; \cref{tab:participant_list}), ML decoding of spared EMG enabled trajectory tracking on a screen for up to three distinct ankle movements (\cref{fig:emg_vs_kinematics}A, B and \cref{fig:task_separation}B-D). Additionally, three participants (S1, S2 and S5) achieved two distinct levels of ankle activation with good accuracy during the target reaching task ($>$ 70\%; \cref{fig:target_reaching}A, B, D; \cref{tab:sup_target_reaching}). Finally, two participants (S1 and S2) used decoded activity to control an electrical stimulator in closed-loop, resulting in more coordinated foot movement (\cref{fig:electrical_stimulation}D) and increased RoM (p $<$ 0.001; \cref{fig:electrical_stimulation}E).

    Our results indicate that residual ankle kinematics after injury can be limited (\cref{fig:emg_vs_kinematics}A, B) and may not always reflect intended movement patterns (\cref{fig:electrical_stimulation}D). These findings align with reports of altered ankle kinematics during gait \cite{krawetz_gait_1996, herrera-valenzuela_derivation_2022, gil-agudo_gait_2011} and suggest that EMG recorded from the target muscle region can provide a measure of spared voluntary foot movement intent (\cref{fig:task_separation}A and \cref{fig:target_reaching}A, B). This approach could therefore complement gait-phase triggers based on motion sensors in commercial drop foot systems \cite{hausdorff_effects_2008, scott_quantification_2013, berenpas_kinematic_2018, embrey_functional_2010}.

    We observed task-modulated neural activity in the affected limb during dorsi-/plantarflexion both through the decomposed MU data (\cref{fig:task_separation}A) and through the ML-decoded movement intent (\cref{fig:task_separation}B). These findings, along with prior reports from our group \cite{oliveira_direct_2024} and others \cite{thomas_motor_1997, mckay_clinical_2004, heald_characterization_2017, ting_sensing_2021}, show that spared voluntary motor commands are still present in the affected limb even after injury. Upper limb studies have also demonstrated that spared neural signals can be leveraged for assistive device control \cite{simpetru_myogestic_2025, osuagwu_active_2020}. Together, these findings support the use of spared EMG from the affected lower limb as a viable control signal for intent-driven neuroprosthetics (\cref{fig:emg_vs_kinematics}B and \cref{fig:electrical_stimulation}B, D, E).

    Previous FES systems have mitigated stimulation artefacts through filtering \cite{osuagwu_active_2020, yeom_autogenic_2010, hambly_comparison_2024} or blanking \cite{minzly_stimulus_1993, de_marchis_multi-contact_2016, thorsen_battery_2009, frigo_emg_2000}, and have typically used a small number of recording electrodes placed over specific muscle sites \cite{osuagwu_active_2020, yeom_autogenic_2010, hambly_comparison_2024, frigo_emg_2000, de_marchis_multi-contact_2016}.
    Our approach used a 32-channel high-density EMG bracelet placed below the knee over muscles relevant to foot movement such as the tibialis anterior and triceps surae (\cref{fig:intro_fig}A and \cref{fig:emg_vs_kinematics}A). This approach could simplify use by recording a broader set of muscles while still capturing residual voluntary motor commands (\cref{fig:task_separation}A). In addition to artefact blanking and filtering, our system integrates a stimulation controller layer with adjustable stimulation timings to reduce artefacts from the ML-decoded EMG signals. Taken together, these components extend prior FES frameworks by enabling ML-based decoding of multiple spared ankle movements derived from EMG signals from the affected limb, as well as supporting closed-loop control with participant-specific parameter tuning (\cref{fig:intro_fig}C, D, \cref{fig:electrical_stimulation}A-E and \Supplementary{} \cref{fig:sup_cursor}).

    Individuals with SCI may retain proportionally controllable EMG activity \cite{ferris_robotic_2009, simpetru_identification_2024} that can modulate stimulation, which has been shown to enhance motor output \cite{osuagwu_active_2020, thorsen_enhancement_2002}. In line with these observations, all five participants (S1-S5; \cref{tab:participant_list}) achieved reliable control of multiple ankle movement patterns (\cref{fig:task_separation}C, D). Three participants (S1, S2 and S5) further showed separable partial and full activation levels for dorsi-/plantarflexion with good accuracy ($>$ 70\%; \cref{tab:sup_target_reaching}). Cursor control performance correlated strongly with the JSD over rest, partial and full EMG activation (r $>$ 0.8; \cref{fig:target_reaching}C), suggesting that cursor feedback can provide a reliable visual indicator of proportional EMG activity.

    One limitation of this study is that stimulator control was not evaluated during gait. This is mainly due to the bulky benchtop and cable-connected stimulator device used ($\sim$2.1 kg; see Methods), which we selected for its wide stimulation parameter range that allows rapid identification of user-specific settings. In addition, some participants (S3, S4 and S5; \cref{tab:participant_list}) were within 6 months post-injury and required walking aids, further limiting feasibility of a walking test with a cable-connected stimulator. Future work will interface our cursor-based controller with a portable stimulator to enable closed-loop control during walking.

    A further limitation is that feedback about decoded movement intensity was primarily visual through the cursor interface. This may increase cognitive load and limit responsiveness compared with other feedback modalities such as haptic feedback. Because sensory preservation varies across SCI and ankle proprioception can be impaired even in motor-incomplete SCI \cite{dambreville_ankle_2019}, an effective feedback strategy may need to be individualized rather than relying on a single solution for all users. Future work should quantify sensory function and evaluate alternative or multimodal feedback (for example, haptic or auditory cues) to reduce reliance on vision and improve functional usability.

    In this study, we validated our cursor control signal for closed-loop control in two participants (S1-S2; \cref{tab:participant_list}), both with chronic injuries of $>$ 4 years and with prior experience with FES-based orthoses. Our goal was to identify a stable myocontrol approach for controlling electrical stimulation in closed-loop. We selected participants whose prior experience with FES orthoses for drop foot correction would indicate a functionally relevant motor response to electrical stimulation.
    While our system demonstrated robust performance for dorsiflexion assistance and potentially for drop foot support, plantarflexion stimulation showed more variable responses. S1 achieved a modest increase in plantarflexion RoM (\cref{fig:electrical_stimulation}E), although baseline motor output was already elevated (MRC grade 3; \cref{tab:participant_list}). In contrast, S2 showed no observable plantarflexion response during stimulation throughout the calibration protocol. Given the low baseline motor output for S2 during plantarflexion (MRC grade 1; \cref{tab:participant_list}) and reliance on walking aids in everyday life, there may be denervation of the plantarflexion muscles that may require alternative stimulation protocols or rehabilitation approaches to restore motor function.
    We acknowledge that these results may not generalize to FES-naive users or acute injuries and that future work is required to validate our system on a wider patient population.

    \section{Conclusion}
    In this study, we identified spared task-modulated neural activity in the affected lower limbs of individuals with motor-incomplete SCI. We further demonstrated that this spared neural activity can be recorded through a wearable 32-channel electrode bracelet and decoded for real-time neuroprosthetic control. All five participants with SCI successfully controlled a cursor on a screen to follow trajectories for 3-4 distinct ankle movements, while three participants showed proportional control of dorsi-/plantarflexion by maintaining two distinct EMG activation levels with accuracy $>$ 70\%. The decoded neural signal captured movement intent that exceeded the physical capabilities of the impaired limb, suggesting that residual motor commands are preserved despite motor impairment.

    Two participants with chronic SCI used the decoded neural signals to control FES in closed-loop, achieving significant increases in dorsiflexion RoM during stimulation (33.6\% and 40\% of typical healthy range, p $<$ 0.001). Correlation between affected and unaffected foot kinematics improved from r $<$ 0.3 to r $>$ 0.7 during stimulation, and one participant demonstrated graded stimulation control with six intensity levels for dorsiflexion and four for plantarflexion. These findings contribute to our fundamental understanding of spared motor control after SCI and show that residual neural activity can be reliably decoded even from limbs with very low muscle strength (e.g., MRC of 0 or 1). The integration of wearable EMG sensing with real-time ML decoding provides a scalable approach for intent-driven neuroprosthetics that could complement existing motion-triggered systems and support individuals with drop foot during activities of daily living.

\end{refsection}

\section*{Data and Code Availability}
The code for the cursor interface (version 0.6.0) is available at https://doi.org/10.5281/zenodo.15700021. Individual participant data may only be shared in encrypted form upon reasonable request and subject to ethics committee approval.

\section*{Acknowledgment}
We would like to acknowledge Prof. Dr. Dario Farina for reviewing an earlier version of our manuscript, as well as Mounir Shaib for his help with preparing and managing the electrical stimulation experiments.

Parts of Figs. \ref{fig:intro_fig}A-D, \ref{fig:sup_marker}A-B, \ref{fig:emg_vs_kinematics}A, \ref{fig:electrical_stimulation}A-B, \ref{fig:task_separation}A-C, and \ref{fig:target_reaching}A-B were created in https://BioRender.com, which we gratefully acknowledge.

\section*{Competing interests}
Co-author Dietmar Fey participated as a subject in this study and is a co-investigator on the funding grants supporting this work. All other authors declare no competing interests.

\renewcommand*{\UrlFont}{\rmfamily}
\printbibliography[heading=bibintoc,title={References},section=1]

\clearpage
\begin{refsection}

    \setcounter{section}{0}
    \renewcommand{\thesection}{S\arabic{section}}
    \setcounter{figure}{0}
    \renewcommand{\thefigure}{S\arabic{figure}}
    \setcounter{table}{0}
    \renewcommand{\thetable}{S\arabic{table}}
    \setcounter{equation}{0}
    \renewcommand{\theequation}{S\arabic{equation}}

    \DeclareFieldFormat{labelnumber}{S#1}

    \newsavebox{\suppCursorbox}
    \savebox{\suppCursorbox}{\includegraphics[width=0.975\textwidth]{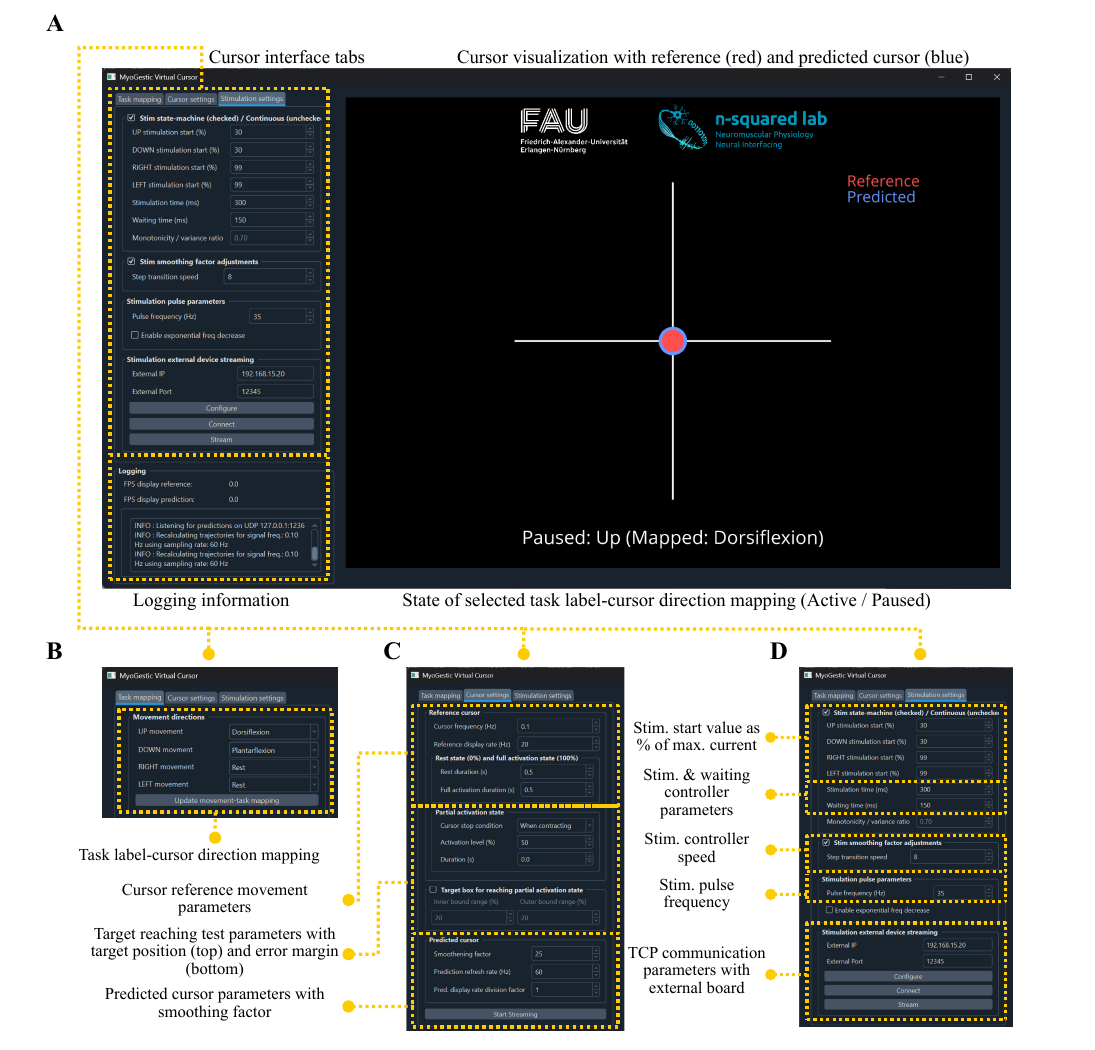}}

    {%
        \makeatletter
        \@dbltoproom=\textheight
        \makeatother
        \twocolumn[%
            \section*{Supplementary Materials}

            \begin{center}
                \usebox{\suppCursorbox}
                \captionof{figure}{\textbf{Cursor interface overview and main functionalities.}
                    \textbf{A.} Main cursor interface window showing the cursor and stimulation parameter tabs (left), and the predicted and reference (right) cursors. Parameters can be adjusted at any point during the experiment to accommodate participant specific movement pace and stimulation requirements. \textbf{B.} Task mapping tab, where the four cursor movement directions can be assigned to any task labels defined in the interface. \textbf{C.} Cursor settings tab, including control parameters for reference and predicted cursor movement, target position and error region used in the target reaching test. \textbf{D.} Stimulation settings tab, including movement specific stimulation start values (mapped to cursor directions), stimulation and waiting state durations, controller speed and pulse frequency. Stimulation parameters are streamed through a Transmission Control Protocol (TCP) communication to a Raspberry Pi board interfacing with the DS8R stimulator and D188 stimulation electrode channel selector.
                }
                \label{fig:sup_cursor}
            \end{center}
            \vspace{1em}
        ]%
    }

    \newsavebox{\suppLDAbox}
    \savebox{\suppLDAbox}{\includegraphics[width=0.975\textwidth]{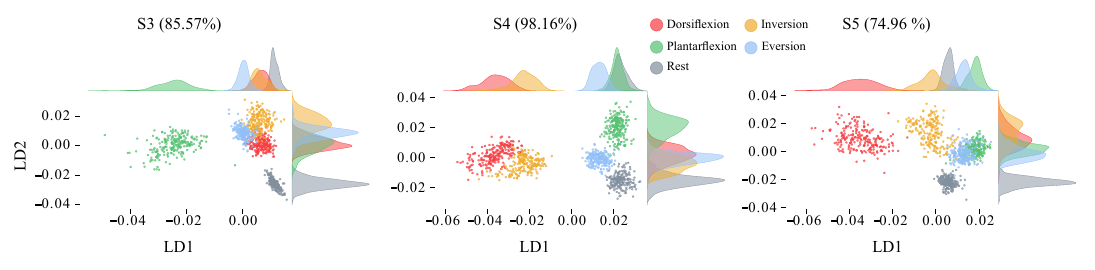}}

    {%
        \makeatletter
        \@dbltoproom=\textheight
        \makeatother
        \twocolumn[%
            \begin{center}
                \captionof{table}{
                    \textbf{Stimulation parameters for each subject.} Values are displayed for the sustained contraction test and, for S1, for the proportional control test (shown in parentheses). The tested models were the Linear Discriminant Analysis (LDA) and gradient boosted decision tree (CatBoost). DF = Dorsiflexion. PF = Plantarflexion. Stim. = stimulation.
                }
                \label{tab:sup_stim_params}

                \rowcolors{2}{gray!15}{white}
                \resizebox{\textwidth}{!}{%
                    \begin{tabular}{c c c c c c c c c}
                        \toprule
                        ID & DF (mA) & PF (mA) & Start thresh. (\%) & Stim. time (s) & Wait time (s) & Freq. (Hz) & Controller speed & Model    \\
                        \midrule
                        S1 & 53 (55) & 42 (42) & 50 (10)            & 1.2 (0.5)      & 0.5 (0.3)     & 25 (35)    & 8                & CatBoost \\
                        S2 & 90      & N/A     & 50                 & 1.2            & 0.7           & 15         & N/A              & LDA      \\
                        \bottomrule
                    \end{tabular}%
                }
            \end{center}
            \vspace{1em}

            \begin{center}
                \usebox{\suppLDAbox}
                \captionof{figure}{\textbf{Linear Discriminant Analysis (LDA) visualization.}
                    The first two LDA components with kernel density estimates and achieved offline accuracies (top) are shown for S3-S5.
                }
                \label{fig:sup_lda}
            \end{center}
        ]%
    }

    \clearpage

    \printbibliography[heading=bibintoc,title={Supplementary References}]

\end{refsection}

\end{document}